\documentclass[sigconf]{acmart}
\usepackage{multirow}
\usepackage{bm}
\usepackage{caption}
\usepackage{subcaption}
\usepackage{tabularx}
\DeclareUnicodeCharacter{2212}{-}

\AtBeginDocument{%
  \providecommand\BibTeX{{%
    \normalfont B\kern-0.5em{\scshape i\kern-0.25em b}\kern-0.8em\TeX}}}

\copyrightyear{2022}
\acmYear{2022}
\setcopyright{rightsretained}
\acmConference[KDD '22]{Proceedings of the 28th ACM SIGKDD
Conference on Knowledge Discovery and Data Mining}{August 14--18,
2022}{Washington, DC, USA}
\acmBooktitle{Proceedings of the 28th ACM SIGKDD Conference on
Knowledge Discovery and Data Mining (KDD '22), August 14--18, 2022,
Washington, DC, USA}
\acmDOI{10.1145/3534678.3539362}
\acmISBN{978-1-4503-9385-0/22/08}

\begin{document}

\title{Using Graph Convolutional Networks to Address fMRI Small Data Problems}


\author{Thomas Screven}
\orcid{0009-0008-1710-3088}
\email{tsscreven@ucdavis.edu}
\affiliation{%
  \institution{University of California, Davis}
  \city{Davis, California}
  \country{USA}}

\author{András Necz}
\email{anecz@ucdavis.edu } 
\affiliation{%
  \institution{University of California, Davis}
  \city{Davis, California}
  \country{USA}}

\author{Jason Smucny}
\orcid{0000-0001-5656-7987}
\email{jsmucny@ucdavis.edu}
\affiliation{%
  \institution{University of California, Davis}
  \city{Davis, California}
  \country{USA}}
\author{Ian Davidson}
\orcid{0000-0002-6481-8018}
\email{davidson@cs.ucdavis.edu}
\affiliation{%
  \institution{University of California, Davis}
  \city{Davis, California}
  \country{USA}}




\renewcommand{\shortauthors}{Trovato and Tobin, et al.}

\begin{abstract}
Although great advances in the analysis of neuroimaging data have been made, a major challenge is a lack of training data. This is less problematic in tasks such as diagnosis, where much data exists, but particularly prevalent in harder problems such as predicting treatment responses (prognosis), where data is focused and hence limited. Here, we address the learning from small data problems for medical imaging using graph neural networks. This is particularly challenging as the information about the patients is themselves graphs (regions of interest connectivity graphs). We show how a spectral representation of the connectivity data allows for efficient propagation that can yield approximately 12\% improvement over traditional deep learning methods using the exact same data. We show that our method's superior performance is due to a data smoothing result that can be measured by closing the number of triangle inequalities and thereby satisfying transitivity. 
\end{abstract}

\ccsdesc[500]{Applied computing~Imaging}
\ccsdesc[500]{Computing methodologies~Learning settings}

\keywords{medical imaging, task fMRI, graph neural networks}

\maketitle

\section{Introduction}

\begin{figure}[htbp]
  \centering
  \includegraphics[width=0.45\textwidth]{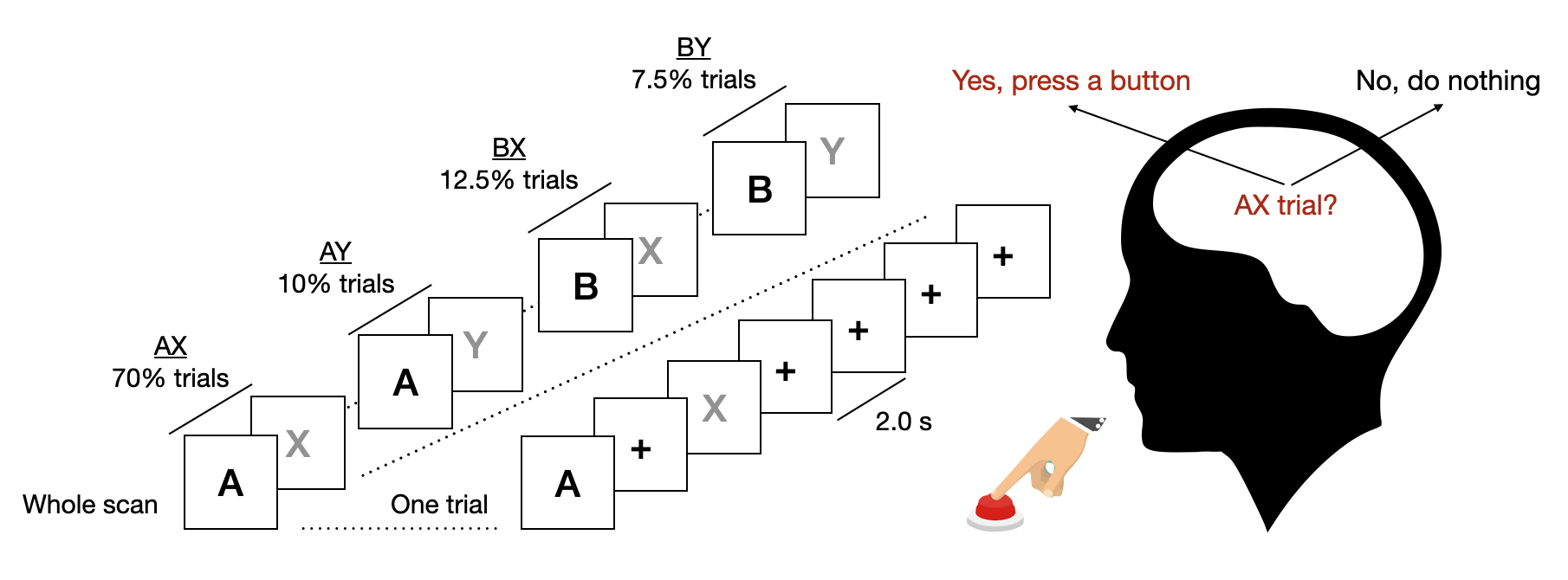}
  \caption{\textbf{An illustration of the AX-CPT task used in our work that generates multiple views of the human brain.} Each trial is started by a cue (`A' or `B') and followed by some \texttt{Rest} frames (`+') and then a probe (`X' or `Y'). The subject is expected to press a button only for the combination where a \texttt{CueA} is followed by a \texttt{ProbeX}. This task elicits an executive reasoning network in the brain. There are 4 types of trials (\texttt{CueA$\rightarrow$ProbeX}, \texttt{CueA$\rightarrow$ProbeY}, \texttt{CueB$\rightarrow$ProbeX}, \texttt{CueB$\rightarrow$ProbeY}). Each type of trial repeats for a varying number of times across subjects.}
  \label{fig:AX-CPT}
\end{figure}


Analysis of functional magnetic resonance imaging (fMRI) data is most frequently performed for patients in ``resting state" (absence of a task) during which the default mode network (DMN) \cite{Greicius2002FunctionalCI} is the most active network. This is useful for \underline{diagnosis} problems such as Post-traumatic stress disorder (PTSD) \cite{https://doi.org/10.1002/da.22481}, Alzheimer's Disease \cite{KOCH2012466} and even Traumatic Brain Injury (TBI) \cite{Bonnelle2012SalienceNI}. Large collections of resting state data exist for many diseases, such as Alzheimer's, with the popular ADNI data set containing thousands of subjects  \cite{jack2008alzheimer}.

However, important problems such as \underline{prognosis}  (the forecast of which subjects will improve (or not) with a specific treatment) can not be easily determined by analyzing readily available resting state fMRI (rs-fMRI) and the DMN \cite{Greicius2002FunctionalCI}. Instead, task fMRI (t-fMRI) data is used when the subject is performing a multi-event task inside the scanner where the task is typically related to the treatment. This limits the data available to learn from as it is typically clinical trial-specific. The application focus of our work is to forecast the symptomatic improvement due to treatment for recent-onset schizophrenia in children by analyzing the baseline AX-CPT (\autoref{fig:AX-CPT}) t-fMRI data \underline{before} the treatment is applied. 
In our work, we have just 82 participants with early psychosis, of which 47 showed at least 20\% improvement in total Brief Psychiatric Rating Scale (BPRS) score between baseline and 12-month follow-up.

\begin{figure*}[htbp]
    \centering
    \includegraphics[width=\textwidth, height=0.3\textheight]{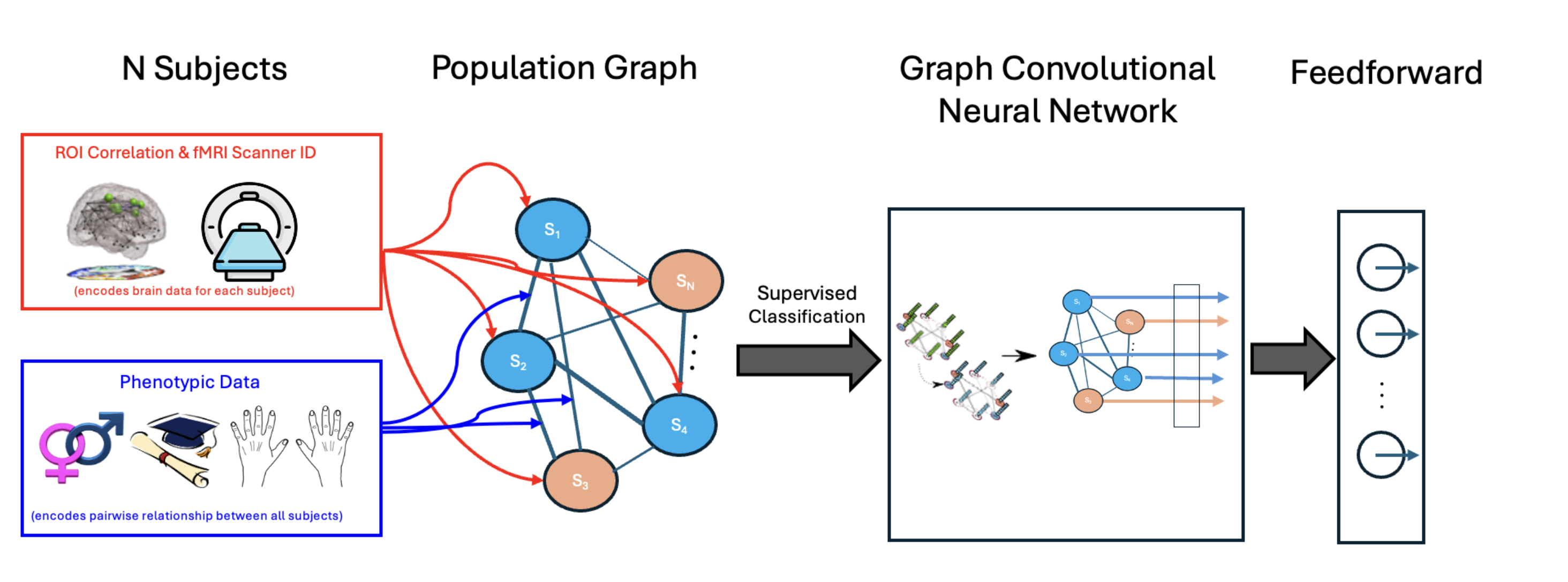}
    \vspace{-10pt} 
    \caption{\textbf{An overview of our model's architecture.}}
    \label{fig:model_overview}
\end{figure*}

Analyzing t-fMRI data is challenging not just due to the small size of the data but also because the underlying computation to perform the prediction is harder. Where as in diagnosis problems, bio-markers are clear and abundant (i.e. the general breakdown of connectivity between \underline{multiple} regions of interests (ROIs) in the brain) for Alzheimer's \cite{jack2008alzheimer} such an obvious biomarker is unlikely to be the situation for prognosis. In our work, we propose a deep graph convolutional learning architecture as shown in \autoref{fig:model_overview} to address the challenges in t-fMRI small data (see \autoref{sec:2}).  
\footnotetext{No instances from the same subject exist in both the validation set and training set.}

The main contributions of this work are as follows:
\begin{itemize}
    \item We explore beyond diagnosis to prognosis (who will respond and not respond to treatment) applications by analyzing t-fMRI data in a novel graph convolutional setting. We show our method (\autoref{sec:3}) can be used to address prognosis problems in a small data situation.
    \item We propose a novel representational scheme for each patient by using a spectral decomposition of their underlying brain connectivity activity. Though spectral analysis of \underline{individual} subject imaging has been explored before \cite{wang2014constrained}, its use as a \underline{representation} scheme is understudied to our knowledge.
    \item We empirically show that our GNN method reaches the subject-wise accuracy of $72.2\% \pm 0.7\% $ in comparison to a regular NN $60.1\% \pm 0.8\% $ despite using the very same data. 
     \item Finally and importantly, we experimentally show that the improved performance of the GNN is because it smooths the data successfully by better removing triangle inequality violations. Closing triangles in the connectivity data allows for a better spectral embedding as spectral methods assume transitivity is satisfied (see Table \ref{tab:noise}).
\end{itemize}
\footnotetext{The data used in (\cite{https://doi.org/10.1002/hbm.25286}) are collected from two scanners with different lengths of trials. 
}

The rest of this paper is organized as follows. 
In \autoref{sec:2}, we explain the t-fMRI data we use, which motivated the graph convolutional learning setting. We explored the deep model architecture and technical details in \autoref{sec:3}. Then, we show our experimental setup and results in \autoref{sec:4}. Finally, a discussion of  related work in  \autoref{sec:6} after which we conclude in \autoref{sec:7}.

\section{Data Setting}\label{sec:2}
In this section, we use the following terminology: subject, scan, trial type, trial, event, and frame. The first two sub-sections are provided for completeness and can be skipped on the first reading. The last two sub-sections provide details required to better understand the representation and learning challenges.

A subject is a participant performing the tasks while the brain activity is recorded. Each subject has a label (responds to treatment or not). A scan is the whole fMRI sequence of one subject. A trial is a snippet of a scan from the beginning of a cue to the last frame before the next cue. An event is either a "Cue", or a "Probe" with "Rest" frames (delay time) between any cues and probes. A frame is a 3-dimensional (3D) picture of the brain consisting of voxels. In our data, BOLD (blood-oxygen-level- dependent) measurements are taken at the voxel level.

\subsection{Data Sample}\label{sec:2.1}

The data sample consisted of 82 individuals with recent onset (<2 years) psychotic disorders. Treatment in the clinic follows a coordinated specialty care (CSC) for early psychosis model delivered by an interdisciplinary treatment team. Treatment includes detailed clinical assessments using gold-standard structured clinical interviews and medical evaluations, targeted pharmacological treatments, including low-dose atypical antipsychotic treatment, individual and family-based psychosocial education and support, cognitive behavioral therapy for psychosis, and support for education and employment. The Structured Clinical Interview for DSM-IV-TR (SCID) (9) was used for diagnosis of psychopathology. Diagnoses were confirmed by a group of trained clinicians during case conferences. All patients reported psychosis onset within two years of the date of informed consent. Patients were excluded for a diagnosis of major medical or neurological illness, head trauma, substance abuse in the previous three months (as well as a positive urinalysis on the day of scanning), Weschler Abbreviated Scale of Intelligence-2 score (WASI-2) (10) score < 70, and magnetic resonance imaging (MRI) exclusion criteria (e.g., claustrophobia, metal in the body). Control participants were excluded for all the above, as well as a history of Axis I mental illness or first-degree family history of psychosis. All participants provided written informed consent in accordance with the Declaration of Helsinki and were compensated for participation. The UCD Institutional Review Board approved the study. Symptoms were assessed using the 24-point Brief Psychiatric Rating Scale (BPRS) (11) rescaled to the lowest score of zero (i.e. score of 24 = score of 0). At baseline, all patients had BPRS scores >= 5 to ensure sufficient resolution to detect a 20\% improvement in score at follow-up.

For the 82 participants, both AX-CPT fMRI data (detailed in \autoref{sec:2.2} and \autoref{sec:2.3}) and clinical phenotypic data were collected. The variables used in this paper are detailed in \autoref{tab:variables}.

\begin{table*}[h]
    \centering
    \caption{\textbf{Phenotype variables used to construct population-level graph between subjects used in GNN.}} 
    \label{tab:variables}
    \begin{minipage}{0.49\textwidth}
        \centering
        \textbf{Quantitative Variables} \\[2pt]
        \begin{tabular}{lccc}
            \toprule    
            Variable & Mean & SD & Sample Size \\
            \midrule
            Education Level$^a$ & 12.8 & 1.78 & 82 \\
            Education Loss vs. Parents$^b$ & 1.63 & 2.87 & 82 \\    
            Crime$^c$ & 4.19 & 2.68 & 82 \\
            Baseline BPRS & 42.7 & 9.63 & 82 \\
            Age & 21.0 & 3.18 & 82 \\
            \bottomrule
        \end{tabular}
        \footnotetext[1]{Years of education.}
        \footnotetext[2]{Years of education less than parents.}
        \footnotetext[3]{Crime level in subject's zip code, measured in violent crimes per thousand.}
    \end{minipage}%
    \hfill
    \begin{minipage}{0.49\textwidth}
        \centering
        \textbf{Qualitative Variables} \\[2pt]
        \begin{tabular}{lcc}
            \toprule    
            Variable & Majority Class and Size & Sample Size \\
            \midrule
            Sex$^a$ & 59 Males & 82 \\
            Handedness$^a$ & 75 Right-Handed & 82 \\
            Diagnosis$^a$$^b$ & 65 Schizophrenia & 82 \\
            Treatment Improvement$^a$ & 47 Improvers & 82 \\
            Race$^c$ & 59 White & 82 \\
            \bottomrule
        \end{tabular}
        \footnotetext[1]{Binary variable, the minority class size can be inferred as $82 - N$.}
        \footnotetext[2]{Minority class is Type I Bipolar Disorder.}
        \footnotetext[3]{Variable is not binary: 59 White, 9 African American/Black, 10 Asian American, 2 Native Hawaiian/Pacific Islander, 1 American Indian/Alaskan Native, 1 multiracial.}
    \end{minipage}
\end{table*}

\subsection{AX-CPT fMRI data}\label{sec:2.2}
Whole brain, single subject fMRI connectivity data were extracted from the AX-CPT using an atlas of 5 mm radius ROIs centered at MNI coordinate locations provided by an fMRI meta-analysis by Power et al. (2011) (16) using the CONN v.21 toolbox (17). Frames with greater than 0.5 mm of movement between them were excluded. Rigid-body movement parameters (x, y, z, roll, pitch, yaw) were used as nuisance regressors when calculating connectivity values. Counts of included frames for each trial-type were AX Tri- als: Mean = 328, S.D. = 96; AY Trials: Mean = 46, S.D. = 15; BX Trials: Mean = 61, S.D. = 17; BY Trials Mean = 41, S.D. = 9. Scanner field strength (1.5T or 3T) was included as a feature. Connectivity values were converted to absolute values before being used in models.

\subsection{Task Description}\label{sec:2.3}
The AX-CPT and associated task parameters have been described in detail elsewhere (3, 12-15). Briefly, participants are presented with a series of cues and probes and are instructed to make a target response (pressing a button with the index finger) to the probe letter "X" only if it is preceded by the cue letter "A." All cues and nontarget probes require nontarget responses (pressing a button with the middle finger). Target sequence trials (i.e., "AX" trials) are frequent (60-70\% occurrence) and set up a prepotent tendency to make a target response when the probe letter X occurs. As a result, a nontarget sequence trial in which any Non-A cue (collectively called "B" cues) is presented and followed by a probe letter X (i.e. "BX" trials) requires proactive cognitive control (e.g. maintenance of the inhibitory rule over the delay time) (13). Consistent with prior work (14), individual subject data was only included in analyses if results suggested the subject understood the AX-CPT (specifically, an accuracy greater than 44\% on AX trials and 50\% on BY trials at both baseline and follow-up). Participants were combined across two task protocols collected from two MRI scanners over a 14-year period. Parameters for each protocol (AX-CPT I and AX-CPT II) are provided in Supplementary Table 1a. The task was presented using EPrime2 software (Psychology Software Tools, Inc.).

\subsection{Small Data and Multi-View Nature}\label{sec:2.4}
This paper is solving a small data problem because the dataset contains only 82 scans from different subjects. At the same time, it is also a multi-view problem. There are \underline{six different trial-types} of the data: CueA, CueB, ProbeAX, ProbeAY, ProbeBX, and ProbeBY. Each trial type reveals different perspectives of brain activity, offering a unique perspective on the subject. Each trial type is treated as a unique view of the subject being scanned, and each view holds a collection of trials. 

\section{Our Approach}\label{sec:3}
We begin by overviewing the entire approach and then going into greater detail in each sub-section:
\begin{itemize}
    \item For each of the six views, we create a different model (see section \ref{sec:3.2}).  Each model is a graph convolutional neural network (GNN). The population structure is derived from the subject's phenotypic data, and the features for each patient/node is their spectrally embedded region of interest (ROI) correlation matrices (see section \ref{sec:3.1}). 
    \item We utilize a majority voting ensemble method to equally consider each trial-type model's predictions. This dynamically combines the multiple views in an instance-specific manner (see section \ref{sec:3.3}).
\end{itemize}

\subsection{Subject Representation and Population Graph Construction}\label{sec:3.1}

\noindent 
We first describe how we represent each subject and then how we construct the graph for the GNN.

\noindent
\textbf{Subject Representation Using Spectral Embedding.} Each subject has a fully connected brain correlation matrix. The correlation matrix identifies co-activation between regions of interest (ROI) in the brain by computing the Pearson correlation between the temporal BOLD (Blood Oxygenation Level Dependent) signals for each ROI. Representing each subject with a full correlation matrix may be suitable for larger data problems but can yield overfitting in small data problems. 
Instead, to find the most active subnetworks in each subject's brain data, we use a spectral embedding approach.

Spectral embedding is performed by converting a subject's correlation matrix into an unnormalized Laplacian matrix, as shown below in equation \autoref{eq1}. Then we find its eigenvectors, which are normalized by their maximum value, meaning all values are between [0,1]. The collection of the largest $k$ (in out experiment $k=10$) eigenvectors serve as the spectral embedding of the patient's correlation matrix. 

Let $C$ be a subject's correlation matrix, $D$ be the degree vector, $L$ be the un-normalized Laplacian matrix, $U$ be the matrix of eigenvectors, and $\lambda$ be the diagonal matrix of eigenvalues then we have:

\begin{equation}
\begin{aligned}
    D &= \sum_{j=1}^{n} C_j \\
    L &= D - C \\
    L U &= \lambda U
\end{aligned}
\label{eq1}
\end{equation}

In the context of correlations between ROI's, less variance signifies that those areas of the brain are highly synchronized with each other. Therefore, we use the eigenvectors with the lowest eigenvalues because these imply active sub-networks in the brain.

\smallskip
\noindent 
\textbf{Population Representation Using Subject Similarity.} We created a weighted adjacency matrix to represent the similarity between subjects. The matrix details how similar the two subjects' phenotypic data are. Two similar subjects have a value closer to 1, while two dissimilar subjects have a value closer to zero. 

Let $W$ be the weighted adjacency matrix where $W[i][j]$ represents the similarity between subject $i$ and subject $j$. Let $f$ be a function that returns the phenotypic data for a subject where $f_2(age)$ returns the age of the second subject. Let $N$ be a function that returns a normalizing constant for that phenotype so that different phenotypes can be aggregated over.

\begin{equation} 
W[i] [j] = \prod _{p \in phenotypes}(1 - \frac{|f_i(p) - f_j(p)|}{C(p)})
\label{eq4}
\end{equation}

To obtain the pairwise relationship between two subjects, we use \autoref{eq4} where $W[i][j]$ is initialized to 1 and then multiplied by the normalized dissimilarity between subject $i$ and subject $j$ for each phenotype.

Using the spectral embedding of the correlation matrices and the population level adjacency matrix, we construct a fully connected population graph where the edges represent population similarity. Each node represents an individual, with the node's features being the spectral embedding of the subject's connectivity matrix. The edge weights between nodes $i$ and $j$ are determined by the adjacency matrix entry at $(i,j)$ ($Y[i][j]=W[j][i]$) based on the similarity of the subjects' phenotypes. We use the phenotype features in Table \ref{tab:variables}.

\subsection{Model Architecture}\label{sec:3.2}

The deep learning architecture created for this problem setting is shown in \autoref{fig:model_overview}. We build six Graph Convolutional Neural Network (GNN) models with the same architecture for each view of the fMRI correlation data. Each model operates on a fully connected population graph incorporating the spectrally embedded ROI correlation data and pairwise phenotypic similarity between subjects.

The GNN models perform feature propagation on each node using its neighbors. Because the graph is fully connected, information is shared between all subjects. The edge weights between subjects in the graph, encoded by their phenotypic similarity, determine how much weight each node has in propagation. The node feature matrix, $H$, encodes the internal feature matrices of each node in the graph. $H$ at layer $k+1$ is updated by multiplying the node feature matrix by the adjacency matrix (\autoref{eq5}).
\begin{equation} 
H^{k+1} = WH^{k}
\label{eq5}
\end{equation}

The graph output of the propagation is fed into a fully connected feedforward neural network, producing binary classification labels. The final graph takes these labels. The nodes in the final graph retain the learned embeddings.

\textbf{Training \& Evaluation.}
For training, a cross-entropy loss function and Adam optimizer are used. The model is evaluated using k-fold cross-validation.

\subsection{Combining Models}\label{sec:3.3}
Our method creates six graph convolutional neural networks, which have all been trained on correlation matrices from different tasks. The six different trial types can create different correlation matrices for the same subject. The subjects and phenotypic data are \underline{identical} between models. We consider each model to be an expert and wish to leverage the knowledge that can be derived from the different tasks.

We accomplish this by implementing a majority vote ensemble method. This protocol combines the binary predictions from each GNN model for each subject. Each model's prediction is equally weighted. At least 4 models need to give a positive prediction for the ensemble to classify a subject as being a treatment improver. Subjects without a majority of models agreeing on a positive classification, including a tie, are classified as treatment non-improvers. 

\section{Experiments and results}\label{sec:4}
In this section, we discuss the experimental settings and results, addressing the following questions:

\begin{itemize}
    \item Can our GNN outperform a standard neural network (NN) model that uses the exact same data? (see section \ref{sec:4.2} and \autoref{tab:model-results})
    \item GNNs do not always work well. Are their properties on the underlying population graph that are more conducive to GNN’s better performance? (see section \ref{sec:4.3} and \autoref{tab:graph-structure})
    \item Is the GNN model's better performance due to smoothing data by reducing the number of triangle inequality violations? (see section \ref{sec:4.4} and \autoref{tab:noise})
\end{itemize}

\subsection{Data and Experimental Setup}\label{sec:4.1}

\textbf{Data Collection.}
Functional images were acquired with a gradient-echo T2* Blood Oxygenation
Level Dependent (BOLD) contrast technique. AX-CPT I was performed in a 1.5T scanner (GE Healthcare), and AX-CPT II in a 3.0T scanner (Siemens). fMRI data were preprocessed using SPM8 (Wellcome Dept. of Imaging Neuroscience, London) as described previously (6, 7). Briefly, images were slice-timing corrected, realigned, normalized to the Montreal Neurological Institute (MNI) template using a rigid-body transformation followed by non-linear warping, and smoothed with an 8 mm full-width-half-maximum Gaussian kernel. All individual fMRI runs had less than 4 mm of translational within-run movement, 3 degrees of rotational within-run movement, and .45 mm of average framewise displacement, calculated using the fsl\_motion\_outliers tool. (https://fsl.fmrib.ox.ac.uk/fsl/fslwiki/FSLMotionOutliers). Mean displacement did not differ between Improvers and Non-Improvers (t = 1.42, p = .16). All participants had at least two fMRI runs surviving these criteria. Preprocessing pipelines were identical for AX-CPT I and II.

\textbf{Experimental Setup}
We studied the performance of our 2-layer GNN and compared it to a 1-layer and a 2-layer neural network (NN). These two other models were constructed using PyTorch neural network modules. For the 2-layer network, processed fMRI connectivity matrices and phenotypic data are the input. This means the NN has the \underline{exact same data} as the GNN, but the two networks use it in different ways. The 1-layer network used just fMRI connectivity matrices. The results show that the GNN performed significantly better than the other two models, resulting in \autoref{tab:model-results}. All models use ten-fold cross-validation. For all models, we used a cross-entropy loss function and stochastic gradient descent for the training optimizer. The GNN has a learning rate of $0.1$. Both the 1 and 2-layer NN have a learning rate of $1 \times 10^{-3}$ and weight decay of $0.1$ every 25 epochs. The dropout rate of the 1-layer network is $0.2$ while the 2-layer network's dropout rate is $0.1$.

\subsection{Model Results}\label{sec:4.2}
\begin{table*}
  \caption{\textbf{Model Performance with 95\% Confidence Interval.} Note the 2-Layer NN uses the exact same data as the GNN (fMRI connectivity and phenotypic data) but in a different manner.} 
  \label{tab:model-results}
  \centering
  \begin{tabular}{lcccccc}
    \toprule    
    Models          & Overall Accuracy         & Improvers Accuracy      & Non-Improvers Accuracy   & ROC AUC                     & F1 Score                \\
    \midrule
    1-Layer NN     & 52.1 (50.5 - 53.6)\%    & 33.8 (27.9 - 39.6)\%    & 66.4 (61.8 - 71.0)\%    & 50.1 (48.4 - 51.8)\%    & 60.2 (57.8 - 62.5)\%    \\    
    2-Layer NN     & 60.1 (59.3 - 60.9)\%    & 23.2 (21.5 - 24.9)\%    & 88.9 (87.7 - 90.1)\%    & 56.0 (55.2 - 56.9)\%    & 71.2 (70.7 - 71.8)\%    \\
    2-Layer GNN (ours)             & 72.2 (71.4 - 72.9)\%    & 76.0 (74.4 - 77.6)\%    & 69.2 (67.8 - 70.5)\%    & 72.5 (71.8 - 73.4)\%    & 73.3 (72.5 - 74.2)\%    \\
    \bottomrule
  \end{tabular}
\end{table*}

Performance metrics for all models are shown in Table 2. It is important to note that the GNN had the effect of improving accuracy significantly for improvers whilst keeping the performance for non-improvers similar. This is a clinically important result as such subjects are the best use of clinical resources and also have a benefit for the subjects as treatments can be time-consuming. Interestingly, the 2-Layer NN uses the population level data (used by the GNN to construct the graph), and the fMRI data represented as a spectral embedding yet performed significantly worse. This shows the improved performance of the GNN is due to how it learns from the data, not due to additional data.

\textbf{Statistical Significance of Results.} A significant effect of model type was observed on accuracy (ANOVA Wilks’ Lambda F(2,23) = 283.4, p < .001). Post-hoc tests revealed significant differences in accuracy between the 1 and 2-layer NNs, the 1-layer NN and GNN, and the 2-layer NN and GNN (all p < .001). Significant effects were also observed on accuracy for improvers F(2,23) = 824.8, p < .001, accuracy for non-improvers(F(2,23) = 509.8, p < .001), AUC (F(2,23) = 448.9, p < .001), and F1 score (F(2,23) = 42.7, p < .001). Post-hoc tests revealed significant differences between all pairwise combinations of models for all of these metrics (p < .01) with the exception of accuracy for non-improvers for the 1-layer CNN vs. the GNN (p = .50).

\subsection{GNN Propagation}\label{sec:4.3}
In the previous section we demonstrated the better performance of our method. Here we try to understand when this will occur and in the next section why it occurred.

Traditional deep learning models, such as the 1 and 2-layer NNs in \autoref{tab:model-results}, treat instances as independent entities. However, our GNN model propagates information between subjects using the population graph structure. This effectively rewrites each subject's connectivity data as a linear combination of its most similar neighbor/subjects (including itself, of course) and their most similar subjects and so on as described in section \ref{sec:3.2}. This process leverages complex relationships between subjects and can mitigate the small data problem, but not always. 

Furthermore, the structure of the patient similarity graph allows effective communication within node communities (see Table \ref{tab:graph-structure}). By sharing data between training instances, our model makes informed transformations to each subject node's embedding using phenotypically similar subjects. This propagation process allows our GNN model to effectively identify patterns that traditional architectures may overlook.

\begin{table*}
  \centering
  \caption{\textbf{Graph Structure}}
  \label{tab:graph-structure}
  \begin{tabular}{lccccc}
    \toprule    
    Graph Type & Average Shortest Path & Local Efficiency & Global Efficiency & Average Clustering & Graph Density \\
    \midrule
    Patient Similarity (see Table \ref{tab:variables}) & 1.01 & 0.998 & 0.988 & 0.997 & 0.406 \\
    Random & 1.25 & 0.875 & 0.875 & 0.750 & 0.512 \\
    Lattice & 5.12 & $4.72 \times 10^{-2}$ & 0.267 & $3.73 \times 10^{-2}$ & $6.08 \times 10^{-2}$ \\
    \bottomrule
  \end{tabular}
\end{table*}

\vspace{-0.1in}
\subsection{Data Smoothing Via Reducing Triangle Inequalities Violations}\label{sec:4.4}

\begin{table*}[htbp]
    \centering
    \caption{\textbf{Percentage of Violation of Triangle Inequality Across All Subjects.}} 
    \label{tab:noise}
    \renewcommand{\arraystretch}{1}
    \setlength{\tabcolsep}{11pt} 
    \begin{tabularx}{\textwidth}{lcccccc}
        \toprule
        & \textbf{CueA} & \textbf{CueB} & \textbf{ProbeAX} & \textbf{ProbeAY} & \textbf{ProbeBX} & \textbf{ProbeBY} \\
        \midrule
        Pre-GNN-Propagation  & 78.6\% & 78.6\% & 78.6\% & 78.5\% & 78.5\% & 78.6\% \\
        Post-GNN-Propagation & 49.7\% & 49.6\% & 49.7\% & 49.6\% & 49.6\% & 49.6\% \\
        \midrule
        SD (Pre) & $2.47\times10^{-3}$\% & $2.48\times10^{-3}$\% & $2.47\times10^{-3}$\% & $2.54\times10^{-3}$\% & $2.74\times10^{-3}$\% & $2.47\times10^{-3}$\% \\
        SD (Post) & $2.06\times10^{-4}$\% & $2.52\times10^{-4}$\% & $2.61\times10^{-4}$\% & $1.97\times10^{-4}$\% & $1.89\times10^{-4}$\% & $2.73\times10^{-4}$\% \\
        \bottomrule
    \end{tabularx}
\end{table*}

The GNN method can be viewed as a pre-processing of the data, in our case, the subject connectivity data. Here, we investigate how our GNN model's feature propagation method favorably changed each subject's correlation matrices.

Our subject representations scheme is a spectral embedding that takes the input correlation matrix between ROIs and attempts to map similarly behaving ROIs close together.  Such a representation scheme makes strong assumptions in particular, that the triangle inequality is satisfied. This is so as if ROI $R1$ is highly correlated with $R2$ and $R2$ is highly correlated with $R3$ then it is assumed that $R1$ and $R3$ are correlated due to transitivity. If the data does not yield this result, and $R1$ and $R3$ are not highly correlated, then this creates challenges. In particular, how to embed $R1$ to be close to $R2$ and $R2$ to be closer to $R3$ yet making $R1$ far from $R3$. Triangle inequality violations are precisely what was occurring in our original data, and we empirically demonstrate that the GNN reduces the number of triangle inequality violations by nearly 30\% (see Table \ref{tab:noise}).

We examined every possible triad of ROIs in each subject's correlation matrix for all six views. Every triad is classified as either satisfying the triangle inequality or not as follows. Let $R1$, $R2$, and $R3$ denote these three ROIs. The subject's correlation matrix contains the correlation value of each pair of ROI combinations. According to the triangle inequality, the absolute value of the correlation between $R1$ and $R3$ must be greater than or equal to the sum of the absolute correlations between $R1$ and $R2$ and $R2$ and $R3$. If not, it signals there is noise in the data in the sense that a triangle inequality is violated. 

We counted the portion of ROI triads with triangle inequality violations for every subject in each view before and after propagation in Table \autoref{tab:noise}. For all six views, the percentage decreases by $\approx29\%$, and the standard deviation decreases by a factor of $\approx10$. 

\textbf{Statistical Significance.} A paired (between subjects before and after applying the GNN) t-test yielded a t-statistic of $1340.67$ and a p-value of $8.44\times10^{-178}$. Additionally, the correlation between the rate at which a subject's triangle inequality was broken before propagation and the change in that subject's node feature matrix after propagation = $0.87$, meaning \underline{more} change occurs on subjects with more triangle inequalities. All of this indicates that our GNN model's success can be partially attributed to reducing noise in the dataset.

\section{Related Work}\label{sec:6}

\textbf{The Deep Learning Studies on fMRI Data.}\label{sec:6.1} Medical imaging analysis has seen considerable development over the last several decades. Thanks to the rapid progress, particularly convolutional neural networks (CNNs) \cite{DL2015}, towards medical imaging analysis \cite{WEN2020101694}. Impressive performance comparable to human experts on image classification, object detection, segmentation, registration, and other tasks \cite{LITJENS201760} has occurred. As one of the most popular modalities, most of the previous works are on resting-state fMRI (rs-fMRI) data. \cite{sarraf2016classification} used convolutional neural networks to classify Alzheimer's brain from the normal healthy brain. \cite{bai2017unsupervised} proposed an unsupervised matrix tri-factorization to discover an underlying network that consists of cohesive spatial regions (nodes) and relationships between those regions (edges) for brain imaging data. Such works on rs-fMRI focus on exploring the intrinsically functionally segregation or specialization of brain regions/networks \cite{logotjetis-fMRI} but are limited on identifying spatiotemporal brain patterns that are functionally involved in specific task performance. 

There exist some  work on using GNN for fMRI data \cite{10.1007/978-3-030-36683-4_65,Parisot2018DiseasePU} but it differs from our work in several important ways. Firstly, previous work is for diagnosis not prognosis, it is for resting state data  not task fMRI data and does not use the spectral embedding representation as we do. Perhaps most importantly it is for larger data sets with the later work \cite{Parisot2018DiseasePU} using the ADNI data set \cite{jack2008alzheimer} which contains thousands of instances not under one hundred like our work.

\textbf{The t-fMRI Studies.}\label{sec:6.2} Recently, the t-fMRI analysis is attracting more and more attention for its ability to connect human activities to brain functioning. In the work of \cite{schwartz2019inducing}, the subjects in the study are asked to read a chapter from a novel while the fMRI scans recording their brain activities are conducted. They fine-tuned a pre-trained BERT model to map the natural language to brain fMRIs. \cite{rieck2020uncovering} used time-varying persistence diagrams to represent the human brain activities when the subjects are watching the movie. \cite{10.1371/journal.pcbi.1006633} studies deep image reconstruction by decoding fMRI into the hierarchical features of a pre-trained deep neural network (DNN) for the same input image. The studies in schizophrenia diagnosis utilizing cognitive control tasks suffered from either small sample size or modest classification performance \cite{pmid29622496}. All these t-fMRI settings are different from the AX-CPT setting, for they don't have multiple types of repeated independent clinical trials to result in one combined evaluation. Instead, their tasks are sequence-to-sequence, guided by the inputs such as series of images and natural languages.

\textbf{The AX-CPT t-fMRI Studies.}\label{sec:6.3} The AX-CPT task is a clinical test on reactive and proactive control processes to identify human cognitive control deficits \cite{LESH2013590}. With modest classification accuracy, the first schizophrenia diagnosis study \cite{yoon2012automated} on the fMRI scans conducted while the cohort subjects completed the AX-CPT task suggests an application to discriminate disorganization level among the patients. \cite{doi:10.1176/appi.ajp.2019.18101126} began the studies on the prognosis of treatment of schizophrenia by analyzing the task-fMRI data. The task-fMRI scans of 82 subjects with psychotic disorders were collected and small regions of interest (ROI) were extracted from the scans for the study. The following work of \cite{https://doi.org/10.1002/hbm.25286} compared machine and naive deep learning-based algorithms for the prediction of clinical improvement in psychosis with the same tast-fMRI data. It achieved ROI voxelwise accuracy of $62.4\%$ using a logistic regression model and $72.6\%$ using a multilayer perceptron model which we used as the baseline to our work. These works highly rely on hand-crafted regions of interest segmentation and they are also analyzing the task-fMRI data on the average activation of some selected keyframes in the scans, which may contribute to a great amount of information loss. In our work, we use the same source of data \footnotemark as the two above works (\cite{doi:10.1176/appi.ajp.2019.18101126, https://doi.org/10.1002/hbm.25286, smucny2021comparing}) on prognosis but only the 51 scans in the 1st protocol (EP1) are included. Different from the above works, we don't need any handcrafted ROI segmentation and the model is working voxelwise on the full brain scans. 

\footnotetext{The data is freely available after requests but can not be publicly posted due to privacy concerns.}

\textbf{The Multi-view Learning and Multi-instance Learning.}\label{sec:6.4} Multi-view learning (\cite{yan2021deep}) and multi-instance learning (\cite{carbonneau2018multiple}) are prevalent in practice; for example, the text content of the web page and the links to the web page are two views of the web page; the gene sub-sequences can be seen as the multiple instances in a bag of the chromosome. Our approach fits the general multi-view multi-instance learning definition but still shows explicit differences from the other latest works in this scope. \cite{nguyen2014labeling, li2017multi, wang2020differentiating, cen2019representation} are non-deep matrix factorization methods or graph representation methods which are not applicable in very high dimensional feature space. In contrast to our multi-view multi-instance setting where a bag of instances represents a view of an example, \cite{yuan2018multi, yang2021deep, xing2019multi} study multi-instance learning on a bag of instances where each instance has multiple views. All these subtle differences invalidate their approaches to be used in our setting. 

\section{Conclusion}\label{sec:7}
The analysis of medical imaging has made great progress, particularly in problems of diagnosis. These results are typically in domains where there are many training instances, as it is possible to aggregate results across different sites, producing data sets with thousands of instances. 

However, a growing important area is prognosis where we attempt to predict whether a subject will respond to at specific treatment based on brain imaging. For such prediction problems there is typically limited data, often numbering in the hundreds. In our experimental setting of predicting treatment response for children with Schizophrenia, we have just 82 instances.

For such small data problems, traditional deep learning methods (see Table \ref{tab:model-results}) produce sub-standard results often because not all subject data is informative. We address the small data challenge with two innovations. Firstly, we showed a spectral embedding of the subject connectivity data produces a simpler representation that prevents over-fitting. Rather than representing each subject with $n^2$ correlations, each subject is represented by $kn$ values where $k=10$ in our experiments. However, a spectral embedding requires well-behaved data in particular that the number of triangle inequality violations is limited. Our second innovation is we use a graph convolution using the subject similarity to smooth out data, and we empirically show that it greatly reduces the number of triangle inequality violations (see Table \ref{tab:noise}). 



\bibliographystyle{ACM-Reference-Format}
\bibliography{KDD-2022}










\end{document}